\documentclass[sigconf]{acmart}
\AtBeginDocument{%
  }

\acmConference[MSR '26]{23rd International Conference on Mining Software Repositories}{April 13--14, 2026}{Rio de Janeiro, Brazil}
\acmBooktitle{23rd International Conference on Mining Software Repositories (MSR '26), April 13--14, 2026, Rio de Janeiro, Brazil}
\setcopyright{none}
\usepackage{booktabs}
\usepackage{multirow}
\usepackage{enumitem}
\usepackage{xspace}

\newcommand{\aidev}{\textsc{AIDev}\xspace}

\newcommand{\codeagent}{{Coding Agent}\xspace}
\newcommand{\codeagents}{{Coding Agents}\xspace}
\newcommand{\codex}{\textsc{OpenAI Codex}\xspace}
\newcommand{\copilot}{\textsc{GitHub Copilot}\xspace}
\newcommand{\devin}{\textsc{Devin}\xspace}
\newcommand{\cursor}{\textsc{Cursor}\xspace}
\newcommand{\claude}{\textsc{Claude Code}\xspace}

\newcommand{\agentpr}{Agentic-PR\xspace}
\newcommand{\agentprs}{Agentic-PRs\xspace}

\newcommand{\humanprs}{Human-PRs\xspace}

\begin{document}

\title{AIDev: Studying AI Coding Agents on GitHub}

\author{Hao Li}
\orcid{0000-0003-4468-5972}
\affiliation{%
  \institution{Queen's University}
  \city{Kingston}
  \state{ON}
  \country{Canada}
}
\email{hao.li@queensu.ca}

\author{Haoxiang Zhang}
\orcid{0000-0002-3921-1724}
\affiliation{%
  \institution{Queen's University}
  \city{Kingston}
  \state{ON}
  \country{Canada}
}
\email{haoxiang.zhang@queensu.ca}

\author{Ahmed E. Hassan}
\orcid{0000-0001-7749-5513}
\affiliation{%
  \institution{Queen's University}
  \city{Kingston}
  \state{ON}
  \country{Canada}
}
\email{ahmed@cs.queensu.ca}

\renewcommand{\shortauthors}{Li et al.}

\begin{abstract}
AI coding agents are rapidly transforming software engineering by performing tasks such as feature development, debugging, and testing. Despite their growing impact, the research community lacks a comprehensive dataset capturing how these agents are used in real-world projects. To address this gap, we introduce AIDev, a large-scale dataset focused on agent-authored pull requests (Agentic-PRs) in real-world GitHub repositories. AIDev aggregates 932,791 Agentic-PRs produced by five agents: OpenAI Codex, Devin, GitHub Copilot, Cursor, and Claude Code. These PRs span 116,211 repositories and involve 72,189 developers. In addition, AIDev includes a curated subset of 33,596 Agentic-PRs from 2,807 repositories with over 100 stars, providing further information such as comments, reviews, commits, and related issues. This dataset offers a foundation for future research on AI adoption, developer productivity, and human-AI collaboration in the new era of software engineering.
\end{abstract}




\maketitle

\section{High-Level Overview}

\begin{figure*}[t]
  \centering
  \includegraphics[width=\linewidth]{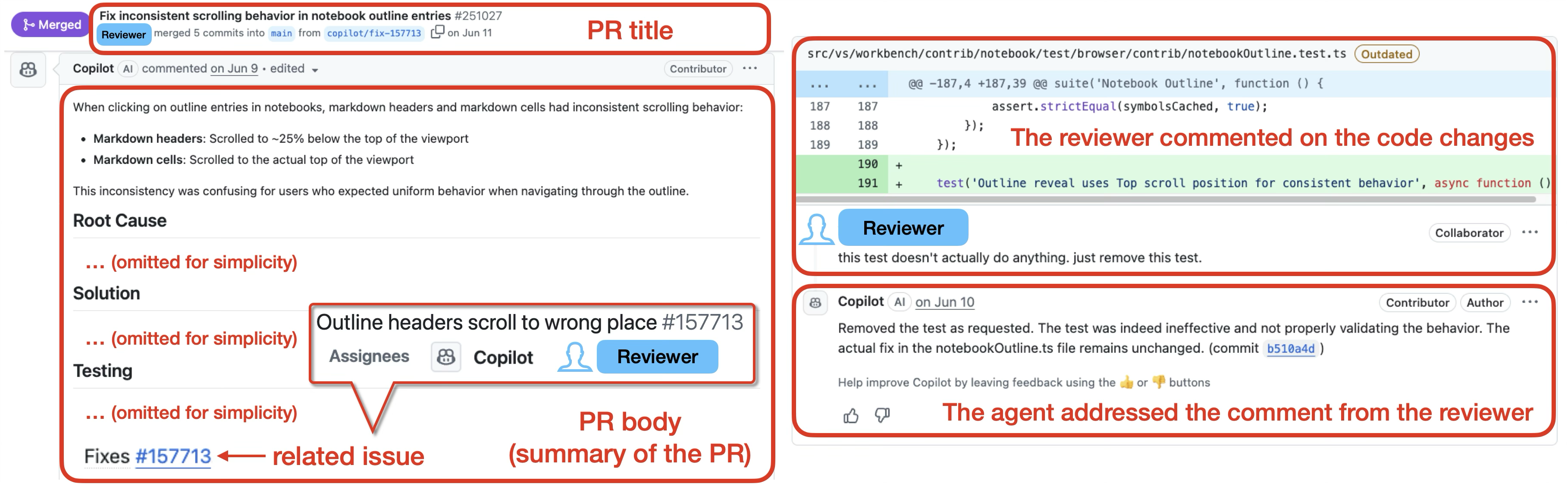}
  \caption{Example of an \agentpr on GitHub. The \codeagent (\copilot) authored a pull request, received feedback from a human reviewer, and addressed the comment in a follow-up commit.}
    \Description[Example of an \agentpr on GitHub]{The \codeagent (\copilot) authored a pull request, received feedback from a human reviewer, and addressed the comment in a follow-up commit.}
  \label{fig:agentic_pr_example}
\end{figure*}

The vision of AI Teammates~\cite{hassan_se3_2024} and recent evidence of their adoption in practice~\cite{li2025riseaiteammatessoftware} signal a major transition in software engineering~(SE). \codeagents are increasingly acting as AI Teammates that participate in core development workflows. They now contribute thousands of pull requests (PRs)\footnote{We refer to PRs authored by \codeagents as \agentprs.} daily, becoming routine actors in collaborative software development. This shift marks the emergence of SE~3.0~\cite{hassan_se3_2024}, where human-AI collaboration is deeply integrated into real-world projects.

Figure~\ref{fig:agentic_pr_example} illustrates how a \codeagent operates within a real GitHub workflow. In this example, the agent (\copilot) is assigned an issue, generates a code patch, and submits a PR with a detailed description. A human reviewer provides feedback, which the agent addresses in a follow-up commit and replies.\footnote{Not all \codeagents currently support addressing review comments.} This interaction showcases the emerging dynamics of human-AI collaboration in software development, where \codeagents not only contribute code but also remain engaged in the review process.

To support systematic study of this paradigm shift, we introduce \aidev, a large-scale dataset of \agentprs from real-world GitHub projects. \aidev comprises 932,791 \agentprs authored by five agents: \codex, \devin, \copilot, \cursor, and \claude, across 116,211 repositories involving 72,189 developers (dataset cutoff: August 1, 2025). Each PR is linked to its corresponding repository and developer, along with additional metadata. For deeper analysis, we curated a subset of 33,596 \agentprs from 2,807 repositories with more than 100 GitHub stars. This enriched subset provides review comments, commit-level diffs, event timelines, and related issues. \aidev enables research on adoption, quality, review, and risks of \codeagents.

\begin{table*}[t]
\centering
\small
\caption{Overview of the \aidev Dataset}
\label{tab:aidev_tables}
\begin{tabular}{llrl}
\toprule
 & \textbf{Table} & \textbf{\# Records} & \textbf{Content} \\ \midrule
\multirow{3}{*}{\textbf{\begin{tabular}[c]{@{}l@{}}Core \\ Metadata\end{tabular}}} & all\_pull\_request & 932,791 & Pull request metadata (title, body, agent, state, timestamps, repository, user) \\
 & all\_repository & 116,211 & Repository-level metadata (name, license, language, URL, stars, forks) \\
 & all\_user & 72,189 & Developer/user metadata (login, followers, creation date) \\ \midrule
\multicolumn{4}{l}{\textit{The following is a subset of PRs from repositories with more than 100 GitHub stars}} \\ \midrule
\multirow{3}{*}{\textbf{\begin{tabular}[c]{@{}l@{}}Core \\ Metadata\end{tabular}}} & pull\_request & 33,596 & Same fields as all\_pull\_request, restricted to curated subset \\
 & repository & 2,807 & Same fields as all\_repository, restricted to curated subset \\
 & user & 1,796 & Same fields as all\_user, restricted to curated subset \\ \midrule
\multirow{3}{*}{\textbf{\begin{tabular}[c]{@{}l@{}}Comments \\ \& Reviews\end{tabular}}} & pr\_comments & 39,122 & Discussion-style comments on PRs (author, body, timestamp) \\
 & pr\_reviews & 28,875 & Review verdicts (e.g., approve or request changes) with metadata (author, body, timestamp) \\
 & pr\_review\_comments & 19,450 & Inline code review comments with file-level context (path, diff hunk, timestamp) \\ \midrule
\multirow{2}{*}{\textbf{\begin{tabular}[c]{@{}l@{}}Commits \\ \& Diffs\end{tabular}}} & pr\_commits & 88,576 & Commits linked to PRs with metadata (SHA, author, message) \\
 & pr\_commit\_details & 711,923 & File-level commit diffs including additions, deletions, and patches \\ \midrule
\multirow{3}{*}{\textbf{\begin{tabular}[c]{@{}l@{}}Issues\\ \& Events\end{tabular}}} 
 & related\_issue & 4,923 & Mapping between PRs and related issues \\
 & issue & 4,614 & GitHub issues related to PRs (title, body, state, user, timestamps) \\
 & pr\_timeline & 325,500 & Full PR event history (e.g., committed, closed, merged, labeled, reviewed) \\ \midrule
\textbf{Annotation} & pr\_task\_type & 33,596 & Automated classification of PR purpose (Conventional Commits categories, GPT-based) \\ \bottomrule
\end{tabular}
\end{table*}

\section{Internal Structure}

Table~\ref{tab:aidev_tables} summarizes the major tables, their sizes, and the types of artifacts they contain. The \aidev dataset provides PR-level, repository-level, and developer-level metadata. The curated subset of repositories with more than 100 GitHub stars provides enriched artifacts such as inline review comments, commit-level diffs, linked issues, and full PR timelines. To support further analysis, the dataset also includes automated annotations of PR purpose (e.g., bug fix, feature, documentation), following the Conventional Commits categories. The complete relational schema is available in our dataset on Hugging Face and Zenodo~(see Section~\ref{sec:links}).

\section{How to Access (Links)}\label{sec:links}

The \aidev dataset is available for download on Hugging Face and Zenodo.
On Hugging Face, the dataset can be explored interactively through the ``Data Studio'' interface, which supports in-browser SQL queries. For reproducibility and ease of use, we also provide example Jupyter notebooks with ready-to-use Google Colab links in our GitHub repository. These notebooks demonstrate how to download, filter, and analyze the dataset.
The related links are provided below:

\begin{itemize}
    \item Hugging Face: \url{https://huggingface.co/datasets/hao-li/AIDev}
    \item Zenodo: \url{https://doi.org/10.5281/zenodo.16899501}
    \item GitHub: \url{https://github.com/SAILResearch/AI_Teammates_in_SE3}
\end{itemize}

\section{Potential Research Questions}

The \aidev dataset opens avenues for a wide range of research opportunities into the role of \codeagents in SE. We outline example research questions below.

\subsection{Adoption and Practices}

\begin{enumerate}
    \item Who adopts \codeagents on GitHub (e.g., newcomers vs. experienced developers)? How do adoption patterns vary across repositories and ecosystems?
    \item What practices (e.g., PR size, task type, and commit granularity) correlate with the quality of \agentprs? How can these practices inform concrete guidelines for developers to work with \agentprs?
    \item How can we identify developers who use \codeagents most effectively? How can we build profiles for these developers? How can we understand their strategies and translate these insights into practical guidance to help others improve their skills and productivity?
\end{enumerate}

\subsection{Code Patch Characteristics}

\begin{enumerate}
    \item How do \agentprs change code (e.g., additions, deletions, files touched)? How consistent are their descriptions with the actual code changes?
    \item To what extent do \agentprs introduce original code versus reusing existing snippets? What are the implications for maintainability?
    \item How do code-change patterns in \agentprs differ from \humanprs (additions vs. deletions, refactorings, file types touched, originality vs. reuse/copying), and what are the implications for software diversity and maintainability?
    \item Do \agentprs conform to project conventions and CI checks (linters, formatters, license headers)? How does conformance relate to review effort and merge outcomes?
\end{enumerate}

\subsection{Testing Behavior}

\begin{enumerate}
    \item How frequently do \codeagents contribute tests? What types (e.g., unit, integration, end-to-end) are most common? What is the test-to-code churn ratio across ecosystems?
    \item When tests are missing in initial \agentprs, do developers intervene to ensure reliable software testing (via follow-up commits or related PRs)?
\end{enumerate}

\subsection{Review Dynamics}

\begin{enumerate}
    \item What aspects of \agentprs (e.g., correctness, style, security, testing) receive the most attention during review?
    \item To what extent do \codeagents address review comments? Which comment types are challenging for agents to resolve?
    \item How can understanding review gaps in AI-generated PRs guide humans in developing new skills for effective collaboration in coding, testing, and reviewing?
\end{enumerate}

\subsection{Failure Patterns and Risks}

\begin{enumerate}
    \item What common failure patterns and code quality issues appear in \agentprs? Why do they occur? How can we leverage these insights to reduce failure rates, optimize human-AI collaboration, and improve AI model training that prioritizes learning from mistakes?
    \item How well can early signals (e.g., PR description, touched paths, and patch characteristics) predict \agentprs rejection or review effort?
    \item How frequently do \agentprs introduce or mitigate security vulnerabilities?
    \item Which \agentprs reach production and persist (i.e., are not reverted or hotfixed)? How do outcomes vary by language, repository maturity, and task type?
    \item What security issues are more prevalent in agent-authored code (e.g., insecure APIs, dependency risks, secrets exposure), and how should review and security practices adapt?
\end{enumerate}

\section{Related Work}\label{sec:related_work}

\subsection{AI Coding Agents}

Recent advances in large language models (LLMs) have enabled coding agents such as SWE-agent~\cite{Yangetal2024}, OpenHands~\cite{Wangetal2024}, and AutoCodeRover~\cite{Zhangetal_12024} that couple LLMs with planning, tool invocation, and repository-level reasoning, enabling tasks such as issue resolution, iterative repair, and test-driven development. Benchmark suites like SWE-bench~\cite{Jimenezetal2023} and its derivatives evaluate whether agents can resolve real GitHub issues end to end, strengthening the link between academic evaluation and production-relevant outcomes.

The technical architectures underlying agentic systems differ from traditional completion-based models. Core designs integrate environment awareness, long-horizon planning, and structured tool use via carefully engineered agent-computer interfaces that expose file I/O, build and test execution, code search, and version-control operations~\cite{Yangetal2024, Maetal2024}. Multi-agent decompositions further segment responsibilities across planning, code editing, navigation, and execution, yielding higher success rates on open-source issue benchmarks than single-agent baselines~\cite{Phanetal2024}. Planning and refinement mechanisms, such as dynamic action resampling, pseudocode-style plan generation, and proposer-ranker loops, improve consistency and robustness across multi-step workflows~\cite{Wenetal2024, Haseeb2025, Wadhwaetal2024}.

Despite rapid progress, important limitations remain. Security analyses have shown elevated rates of vulnerabilities in AI-generated code, raising the stakes as agents obtain write or commit permissions in real repositories~\cite{Pearceetal2021, Kozaketal2025, Sajadietal2025}. Usability studies identify challenges related to latency, context handling, and user trust, particularly when agents act proactively or at repository scale~\cite{Puetal2025, Eibletal2025}. Position papers argue that achieving programming with trust will require constraining autonomy through verifiable analyses, richer testing, and auditable traces of thought, action, and result trajectories~\cite{Bouzeniaetal2025}. Work on RepairAgent and related systems illustrates how structured finite-state workflows and tool gating can align agent behavior with developer expectations while improving reliability~\cite{Bouzeniaetal2024, Wadhwaetal2024}.

Taken together, prior research depicts a field transitioning from token-level assistance toward agentic software engineering~\cite{hassan2025agenticse}, in which LLM-driven systems plan, execute, and iterate over complex development tasks with varying degrees of autonomy~\cite{Jinetal2024, Wangetal2025}. However, much of the existing evidence derives from controlled studies, benchmarks, or small-scale deployments. This gap motivates our \aidev dataset, which captures \agentprs at scale across real GitHub repositories. By linking PR metadata with review artifacts, commit diffs, issue relations, and event timelines, \aidev provides a foundation for studying adoption, quality, review dynamics, and risks of \codeagents in the wild.

\subsection{Automated Contributions in Open Source Software}

Prior work on automation in open source software has examined rule-based tools and development bots that react to well-defined triggers, operate within narrow scopes, and surface changes through templated interactions such as scripted comments or dependency update pull requests~\cite{Shihabetal2022, Erlenhovetal2022}. These systems have helped scale maintenance work, most prominently in dependency management, but they rarely exhibit the kind of initiative, contextual reasoning, or dialogical interaction seen in human contributors~\cite{Erlenhovetal2022, Casseeetal2020}.

Bot-generated PRs tend to have lower acceptance rates, slower interactions, and well-documented issues with noise and notification fatigue, particularly in the context of dependency updates~\cite{Wyrichetal2021, Heetal2022}. These patterns have been linked to limited trust, brittle automation, and interruptions to established workflows~\cite{Wesseletal2021, Erlenhovetal2020}. By contrast, recent observations of LLM use within PR workflows indicate that developers turn to these tools for larger, more complex changes. PRs associated with ChatGPT involvement are substantially slower to close and entail heavier review, suggesting qualitatively different kinds of contributions than routine maintenance~\cite{Chouchenetal2024, Champaetal2024}. For dependency and repair bots, rejection often stems from volume, false alarms, or brittle patches~\cite{Heetal2022, Rombautetal2022, Urlietal2017}. Early evidence suggests that LLM involvement correlates with heavier review workloads and longer time-to-merge, consistent with a shift from maintenance micro-changes to higher-stakes modifications~\cite{Chouchenetal2024, Champaetal2024, Ogenrwotetal2025}.

\subsection{Empirical Studies of AI Adoption and Developer Productivity}

Field and observational studies complement surveys by revealing how developers actually interact with AI tools. In practice, many engineers rely on AI tools more for guidance, exploration, and sensemaking than for drop-in code generation, weaving suggestions into existing validation and repair routines \cite{Khojahetal2024}. Interview-based investigations with early adopters broaden this picture across people, processes, and products, noting both workflow accelerations and socio-technical frictions as teams integrate LLM assistance into the software lifecycle~\cite{Tabarsietal2025}. Laboratory and think-aloud studies sharpen these insights: although participants often prefer working with tools like GitHub Copilot, measurable efficiency gains are inconsistent when tasks require deep comprehension or debugging of suggested code~\cite{Vaithilingametal2022}. Grounded analyses of interaction further distinguish between acceleration, using AI tools to speed routine authoring, and exploration, using them to plan structure or discover APIs \cite{Barkeetal2022}. Recent work also foregrounds the affective dimension of AI tool use, documenting frustration and emotional strain that, while common, rarely lead to abandonment of these tools \cite{Montesetal2025}.

Telemetry and usage analytics offer population-scale evidence about real-world behavior. By linking IDE measurements with self-reports, Ziegler et~al.~\cite{Ziegleretal2022} demonstrate that the acceptance rate of shown suggestions is a stronger predictor of perceived productivity than coarse output metrics, and that acceptance varies markedly across users and over time. Eye-tracking and IDE-instrumented experiments additionally show that awareness of code provenance changes behavior: when developers know a snippet was AI-generated, they search more, validate more, and experience higher cognitive load while nonetheless improving performance on some tasks \cite{Tangetal2024}. Natural experiments on platform-wide rollouts provide convergent evidence for ecosystem-level impact: when Copilot selectively supported certain languages on GitHub, contribution volumes shifted in ways consistent with augmented collaborative innovation for supported languages \cite{Yeverechyahuetal2024}. Complementary analyses map AI use to specific tasks, including bug fixing and testing, suggesting that effectiveness depends on the surrounding validation and review practices into which suggestions are embedded \cite{Xiaetal2022, Sobaniaetal2023, Monteiroetal2023}.

Direct measurements of productivity remain mixed but cautiously positive, reflecting the inherent difficulty of quantifying developer work. Controlled studies frequently find improvements in perceived productivity and quality of starting points, even when completion time or task success rates do not uniformly improve \cite{Vaithilingametal2022, Barkeetal2022}. Broader evaluation frameworks such as RealHumanEval~\cite{Mozannaretal2024} seek to capture multidimensional outcomes across participants, tasks, and assistance modalities. Quasi-experimental evidence from policy shocks likewise indicates measurable gains: for example, a temporary regional restriction on LLM access enabled an estimate of a 6.4\% productivity reduction during the ban, with heterogeneous effects by experience level \cite{Bonabietal2025}. Systematic reviews synthesize these threads, noting that while most studies now adopt multidimensional perspectives, the field still leans heavily on self-reports and short-term, individual-level designs~\cite{Mohamedetal2025}.

\bibliographystyle{ACM-Reference-Format}
\bibliography{main}

\end{document}